\documentclass{article}
\newcommand{\be}{\begin{equation}}
\newcommand{\ee}{\end{equation}}
\newcommand{\bea}{\begin{eqnarray}}
\newcommand{\eea}{\end{eqnarray}}

\newcommand{\ve}{\varepsilon}
\newcommand{\ds}{\displaystyle}

\begin{document}

\title{On bare masses in time-symmetric initial-value solutions for two black
holes}

\author{Piotr Jaranowski
\\{\it Institute of Physics, Bia{\l}ystok University}\\
  {\it Lipowa 41, 15-424 Bia{\l}ystok, Poland}
\thanks{E-mail: pio@alpha.uwb.edu.pl}\vspace{2ex}\\
Gerhard Sch\"afer
\\{\it Theoretisch-Physikalisches Institut, 
       Friedrich-Schiller-Universit\"at}\\
  {\it Max-Wien-Platz 1, 07743 Jena, Germany}
\thanks{E-mail: gos@tpi.uni-jena.de}}

\date{}

\maketitle

\begin{abstract}

The Brill-Lindquist time-symmetric initial-value solution for two uncharged
black holes is rederived using the Hamiltonian constraint equation with Dirac
delta distributions as a source for the binary black-hole field. The bare
masses of the Brill-Lindquist black holes are introduced in a way which is
applied, after straightforward modification, to the Misner-Linquist binary
black-hole solution.

\vspace{0.3cm}\noindent PACS number(s):
04.70.Bw, 04.20.Ex, 04.20.Fy
\end{abstract}
\vspace{0.5cm}

In a recent paper \cite{JS99} by the authors on post-Newtonian expansion of the
Brill-Lindquist (BL) \cite{BL63} and Misner-Lindquist (ML) \cite{M63,L63}
solutions of the time-symmetric initial-value geometry for two uncharged black
holes it turned out that the conformally flat metric coefficients of the BL and
ML solutions coincide up to the third post-Newtonian approximation (so they
begin to deviate at the fourth post-Newtonian order), provided the bare masses
in the both solutions have been identified. By shifting the centers of the ML
black holes the coincidence has been achieved up to the fifth post-Newtonian
order. It is the aim of the present paper to give arguments for the assumed, in
\cite{JS99}, identification of the bare masses. Whereas in the papers
\cite{BL63,M63,L63} in case of uncharged black holes a linear vacuum Einstein
constraint equation has been solved and investigated, in the present paper we
shall apply the non-linear Hamiltonian constraint equation with some Dirac
delta distributions as sources for the binary BL and ML black-hole fields.

We use units in which $16\pi G=c=1$, where $G$ is the Newtonian gravitational
constant and $c$ the velocity of light. In the paper all vectors and their
lengths are defined in the 3-dimensional Euclidean space endowed with a
standard Euclidean metric; ${\bf x}$ is the position of an arbitrary point in
this space.

\vspace{0.2cm}

The time-symmetric initial-value solutions, which we also call (instantenously)
static solutions, solve the Hamiltonian constraint equation under the
assumption that the linear momenta of the black holes, the transverse-traceless
part of the three metric, and the field conjugate momenta are put equal to
zero. The static three metric can be put into conformally flat form
[cf.\ Eq.\ (1) in \cite{JS99}]
\be
\label{metric}
g_{ij} = \left(1+\frac{1}{8}\phi\right)^4\delta_{ij},
\ee
and the static Hamiltonian constraint equation reads
\be
\label{ce}
\left(1+\frac{1}{8}\phi\right)\Delta\phi = {{\cal T}^0}_0,
\ee
where ${{\cal T}^0}_0$ is a 00-component of the matter stress-energy tensor
density.

For the BL solution the function $\phi$ from Eq.\ (\ref{metric}) equals
(cf.\ Ref.\ \cite{BL63} and Appendix A in \cite{AP97})
\be
\label{BL}
\phi^{\rm BL} = 8\left(
\frac{\alpha_1}{|{\bf x}-{\bf x}_1|} + \frac{\alpha_2}{|{\bf x}-{\bf x}_2|}
\right),
\ee
where $\alpha_1$ and $\alpha_2$ are some positive parameters, and ${\bf x}_a$
($a=1,2$) is a position vector of the point representing the $a$th hole.
For the BL solution we employ the following stress-energy tensor density
\be
\label{BLem}
{{\cal T}^0}_0= -\left( {\stackrel{0}m}_1\delta({\bf x}-{\bf x}_1)
+ {\stackrel{0}m}_2\delta({\bf x}-{\bf x}_2) \right),
\ee
where ${\stackrel{0}m}_1$ and ${\stackrel{0}m}_2$ are constant ``naked'' masses
of the black holes.

Let us assume that there exists a solution of the constraint
equation (\ref{ce}) with the source (\ref{BLem}) in the form given by
(\ref{BL}). Plugging (\ref{BL}) and (\ref{BLem}) into Eq.\ (\ref{ce}) one
obtains
\bea
\label{BL1}
&\ds \left( \alpha_1+\frac{\alpha_1^2}{|{\bf x}-{\bf x}_1|}+
\frac{\alpha_1\alpha_2}{|{\bf x}-{\bf x}_2|} \right)\delta({\bf x}-{\bf x}_1)
+ \left( \alpha_2+\frac{\alpha_2^2}{|{\bf x}-{\bf x}_2|}+
\frac{\alpha_1\alpha_2}{|{\bf x}-{\bf x}_1|} \right)\delta({\bf x}-{\bf x}_2)&
\nonumber\\[2ex]
&\ds = \frac{1}{32\pi} \left( {\stackrel{0}m}_1\delta({\bf x}-{\bf x}_2)
+ {\stackrel{0}m}_2\delta({\bf x}-{\bf x}_2) \right).&
\eea
In Eq.\ (\ref{BL1}) we replace $f({\bf x})\delta({\bf x}-{\bf x}_a)$ by
$f_{\rm reg}({\bf x}_a)\delta({\bf x}-{\bf x}_a)$, where the regularized value
of the function $f$ at its (possibly) singular point ${\bf x}={\bf x}_a$ we
define by means of the Hadamard's ``partie finie'' procedure. We expand
$f\left({\bf x}_a+\ve{\bf n}\right)$ (where ${\bf n}$ is a unit vecor) into a
Laurent series around $\ve=0$ and as the regularized value of the function $f$
at ${\bf x}_a$ we take the zero-order coefficient of the series averaged over
all directions ${\bf n}$:
\be
\label{hpf}
f\left({\bf x}_a+\ve{\bf n}\right)
=\sum\limits_{m = -N}^{\infty}a_m({\bf n})\,\ve^m,\quad
f_{\rm reg}\left({\bf x}_a\right):=\frac{1}{4\pi}\oint\!d\Omega\,a_0({\bf n}).
\ee
Using the definition (\ref{hpf}) we obtain
\be
\label{hpfr}
\frac{\delta({\bf x}-{\bf x}_1)}{|{\bf x}-{\bf x}_1|} = 0,\quad
\frac{\delta({\bf x}-{\bf x}_1)}{|{\bf x}-{\bf x}_2|}
= \frac{\delta({\bf x}-{\bf x}_1)}{|{\bf x}_1-{\bf x}_2|}.
\ee
Applying Eqs.\ (\ref{hpfr}) into Eq.\ (\ref{BL1}) and comparing the
coefficients at $\delta({\bf x}-{\bf x}_1)$ and $\delta({\bf x}-{\bf x}_2)$ on
both sides of (\ref{BL1}) we get the relations
\be
\label{BLbma}
\frac{1}{32\pi}m_1 = \alpha_1 + \frac{\alpha_1\alpha_2}{|{\bf x}_1-{\bf x}_2|},
\quad
\frac{1}{32\pi}m_2 = \alpha_2 + \frac{\alpha_1\alpha_2}{|{\bf x}_1-{\bf x}_2|}.
\ee
Here we have replaced the constant naked masses ${\stackrel{0}m}_1$ and
${\stackrel{0}m}_2$ of the black holes by their constant ``bare'' masses $m_1$
and $m_2$. The bare masses take part in the black-hole black-hole interaction
due to the regularization procedure applied to Eq.\ (\ref{BL1}). The
expressions (\ref{BLbma}) coincide with the formulae for the bare masses
obtained by Brill and Lindquist in \cite{BL63} (see Eq.\ (14) in \cite{BL63}).

\vspace{0.2cm}

Now we rederive the bare masses for the BL solution in a way which will be
applicable also for the ML solution. We start from assuming the following
definition of the bare masses:
\be
\label{bm}
m_a := {\stackrel{0}m}_a + \left[
\int\limits_{B\left({\bf x}_a,\ve_a\right)}
\kern-2ex d^3x \frac{1}{8}\phi\Delta\phi
\right]_{{\rm self}(a)},\quad a=1,2,
\ee
where $B\left({\bf x}_a,\ve_a\right)$ is any ball centered at ${\bf x}_a$
with the radius $\ve_a<|{\bf x}_1-{\bf x}_2|$, the subscript ``self$(a)$''
denotes the infinite self-interaction term associated with the $a$th black hole.
We substitue Eqs.\ (\ref{BL}) and (\ref{BLem}) into (\ref{ce}) and integrate
both sides over the ball, say, $B\left({\bf x}_1,\ve_1\right)$. The result
is
\be
\label{BLbm1}
{\stackrel{0}m}_1 = 32\pi \left(\alpha_1
+ \alpha_1^2 \int\limits_{B\left({\bf x}_1,\ve_1\right)}
\kern-2ex  d^3x \frac{\delta({\bf x}-{\bf x}_1)}{|{\bf x}-{\bf x}_1|}
+ \alpha_1\alpha_2 \int\limits_{B\left({\bf x}_1,\ve_1\right)}
\kern-2ex d^3x \frac{\delta({\bf x}-{\bf x}_1)}{|{\bf x}-{\bf x}_2|}
\right).
\ee
Equation (\ref{BLbm1}) yields
\be
\label{BLbm2}
\left[
\int\limits_{B\left({\bf x}_1,\ve_1\right)}
\kern-2ex d^3x \frac{1}{8}\phi\Delta\phi
\right]_{{\rm self}(1)} = -32\pi \alpha_1^2
\int\limits_{B\left({\bf x}_1,\ve_1\right)}
\kern-2ex d^3x \frac{\delta({\bf x}-{\bf x}_1)}{|{\bf x}-{\bf x}_1|}.
\ee
Making use of Eqs.\ (\ref{bm}) and (\ref{BLbm2}), Eq.\ (\ref{BLbm1}) can be
written as
\be
\label{BLbm3}
\frac{1}{32\pi}m_1 = \alpha_1 + \frac{\alpha_1\alpha_2}{|{\bf x}_1-{\bf x}_2|}.
\ee
The derivation of the formula for $m_2$ is analogous. The expression
(\ref{BLbm3}) coincides with the first of the formulae (\ref{BLbma}).

\vspace{0.2cm}

The ML solution is described by the function (cf.\ Refs.\ \cite{M63,L63} and
Appendix B in \cite{AP97})
\be
\label{ML}
\phi^{\rm ML} = 8\sum_{n=1}^{\infty}
\left(\frac{a_n}{|{\bf x}-{\bf d}_n|}+\frac{b_n}{|{\bf x}-{\bf e}_n|} \right),
\ee
where ${\bf d}_1$ is the position of the center of the black hole 1 of radius
$a\equiv a_1$ and ${\bf e}_1$ is the position of the center of the black hole 2
of radius $b\equiv b_1$; ${\bf d}_n$ and ${\bf e}_n$ ($n\ge2$) are the
positions of the image poles of black hole 1 and 2, respectively, $a_n$ and
$b_n$ ($n\ge2$) are the corresponding weights. All weights $a_n$, $b_n$ and
vectors ${\bf d}_n$, ${\bf e}_n$ can be expressed by the radii $a$, $b$ of the
black holes and the vector ${\bf d}_1-{\bf e}_1$ connecting their centers. Note
also that odd images of the black hole 1 and even images of the black hole 2
are located inside the sphere $|{\bf x}-{\bf d}_1|=a$, whereas even images of
the black hole 1 and odd images of the black hole 2 are within the sphere
$|{\bf x}-{\bf e}_1|=b$. For the ML solution we make the following ansatz for
the energy-momentum tensor:
\be
\label{MLem}
{{\cal T}^0}_0= -\sum_{n=1}^\infty \left(
{\stackrel{0}\mu}_n\delta({\bf x}-{\bf d}_n)
+ {\stackrel{0}\nu}_n\delta({\bf x}-{\bf e}_n) \right),
\ee
where ${\stackrel{0}\mu}_n$ and ${\stackrel{0}\nu}_n$ are the constant naked
masses of the image poles located at ${\bf d}_n$ and ${\bf e}_n$, respectively.

The bare mass of the ML black hole collects all naked masses of the image poles
associated with one of the throats together with all self-interaction terms, 
including infinite ones, between these images:
\bea
\label{MLbm1a}
m_1 &:=&
\sum_{n=1}^\infty \left({\stackrel{0}\mu}_{2n-1}+{\stackrel{0}\nu}_{2n}\right)
+ \left[ \int\limits_{B\left({\bf d}_1,\ve_1\right)}
\kern-2ex d^3x \frac{1}{8}\phi\Delta\phi
\right]_{{\rm self}(1)},
\\[2ex]
\label{MLbm1b}
m_2 &:=&
\sum_{n=1}^\infty \left({\stackrel{0}\mu}_{2n}+{\stackrel{0}\nu}_{2n-1}\right)
+ \left[ \int\limits_{B\left({\bf e}_1,\ve_2\right)}
\kern-2ex d^3x \frac{1}{8}\phi\Delta\phi
\right]_{{\rm self}(2)},
\eea 
where the ball $B\left({\bf d}_1,\ve_1\right)$ has the radius $a<\ve_1<c-b$ and
the ball $B\left({\bf e}_1,\ve_2\right)$ has the radius $b<\ve_2<c-a$,
$c:=|{\bf d}_1-{\bf e}_1|$ is the coordinate distance between the centers of
the black holes. We substitute Eqs.\ (\ref{ML}) and (\ref{MLem}) into
(\ref{ce}) and integrate both sides over the ball, say,
$B\left({\bf d}_1,\ve_1\right)$. The result is
\bea
\label{MLbm2}
\ds \sum_{n=1}^\infty
\left({\stackrel{0}\mu}_{2n-1}+{\stackrel{0}\nu}_{2n}\right)
&=& 32\pi \sum_{n=1}^\infty \left(a_{2n-1}+b_{2n}\right)
\nonumber\\[2ex]&&\ds
+ 32\pi \sum_{m=1}^\infty\sum_{n=1}^\infty
\int\limits_{B\left({\bf d}_1,\ve_1\right)}\kern-2ex d^3x
\Bigg[
\left( \frac{a_m a_{2n-1}}{|{\bf x}-{\bf d}_m|}
+ \frac{b_m a_{2n-1}}{|{\bf x}-{\bf e}_m|} \right)
\delta({\bf x}-{\bf d}_{2n-1})
\nonumber\\[2ex]&&\ds
+ \left( \frac{a_m b_{2n}}{|{\bf x}-{\bf d}_m|}
+ \frac{b_m b_{2n}}{|{\bf x}-{\bf e}_m|} \right)
\delta({\bf x}-{\bf e}_{2n}) \Bigg].
\eea
Equation (\ref{MLbm2}) yields
\bea
\label{MLbm3}
\left[
\int\limits_{B\left({\bf d}_1,\ve_1\right)}
\kern-2ex d^3x \frac{1}{8}\phi\Delta\phi
\right]_{{\rm self}(1)}
&=& -32\pi
\sum_{m=1}^\infty\sum_{n=1}^\infty
\int\limits_{B\left({\bf d}_1,\ve_1\right)}\kern-2ex d^3x
\Bigg[
\left( \frac{a_{2m-1} a_{2n-1}}{|{\bf x}-{\bf d}_{2m-1}|}
+ \frac{b_{2m} a_{2n-1}}{|{\bf x}-{\bf e}_{2m}|} \right)
\delta({\bf x}-{\bf d}_{2n-1})
\nonumber\\[2ex]&&\ds
+ \left( \frac{a_{2m-1} b_{2n}}{|{\bf x}-{\bf d}_{2m-1}|}
+ \frac{b_{2m} b_{2n}}{|{\bf x}-{\bf e}_{2m}|} \right)
\delta({\bf x}-{\bf e}_{2n}) \Bigg].
\eea
The terms on the right-hand side of this equation which are infinite
``renormalize'' the constant naked masses ${\stackrel{0}\mu}_n$ and
${\stackrel{0}\nu}_n$ to a kind of constant bare masses, like in Eq.\
(\ref{bm}). However, there are still finite terms left. One can show that their
sums diverge \cite{L63} so that a further renormalization of the whole mass of
the image ensemble is needed, which finally results in the bare mass $m_1$.
This latter infinite renormalization can be used as argument for treating the
mass $m_1$ as constant. Making use of Eqs.\ (\ref{MLbm1a}) and (\ref{MLbm3}),
Eq.\ (\ref{MLbm2}) can be written as
\bea
\label{MLbm4}
\frac{1}{32\pi} m_1 &=& \sum_{m=1}^\infty \bigg[ a_{2m-1} + b_{2m}
\nonumber\\[2ex]&&
+ \sum_{n=1}^\infty \left(
\frac{a_{2m} a_{2n-1}}{|{\bf d}_{2m}-{\bf d}_{2n-1}|}
+ \frac{a_{2m} b_{2n}}{|{\bf d}_{2m}-{\bf e}_{2n}|}
+ \frac{b_{2m-1} a_{2n-1}}{|{\bf e}_{2m-1}-{\bf d}_{2n-1}|}
+ \frac{b_{2m-1} b_{2n}}{|{\bf e}_{2m-1}-{\bf e}_{2n}|} \right)\bigg].
\eea
Analogously we can obtain the formula for the mass $m_2$. Expression
(\ref{MLbm4}) for the bare mass $m_1$ agrees with the formula introduced by
Lindquist in \cite{L63} (cf.\ Eq.\ (B17) in \cite{AP97}).

The comparison of Eq.\ (\ref{bm}) with Eqs.\ (\ref{MLbm1a}) and (\ref{MLbm1b})
shows that  the bare masses are introduced analogously for the BL and ML
solutions. Thus, for the comparison of the both solutions, the identification
of the bare masses seems to be natural, in particular when performing
post-Newtonian expansions where the black holes  are assumed to be far away
from each other.

\bigskip\noindent
{\bf Acknowledgment}

\noindent This work was supported by the Max-Planck-Gesellschaft Grant No.\ 
02160-361-TG74 (GS).

\end{document}